\documentclass[aps,preprint,superscriptaddress]{revtex4}
\newcommand{\Bra}[1]{\left \langle #1 \right |} 
\newcommand{\Ket}[1]{\left | #1 \right \rangle}  

\newcommand{\vtwo}[2]{\left ( \begin{array}{c} #1 \\ #2 \end{array} 
  \right )}
\newcommand{\X} {(\Omega ^+ \Omega)}
\newcommand{\suma}{\sum\limits_{n=1}^{\infty}}
\newcommand{\sumb}{\sum\limits_{i=0}^n}
\newcommand{\be}{\begin{equation}}
\newcommand{\ee}{\end{equation}}
\newcommand{\norm}[1]{\{#1\}}
\begin{document}

\title{Conservation of connectivity of model-space effective
  interactions under a class of similarity transformation}
\author{Chang-Kui Duan}
\affiliation{Institute of Applied Physics and College of
Electronic Engineering, Chongqing University of Posts and 
Telecommunications, Chongqing 400065, China}
\affiliation{Department of Physics and Astronomy, 
University of Canterbury, Christchurch, New Zealand
}
\author{Yungui Gong}
\author{Hui-Ning Dong}
\affiliation{Institute of Applied Physics and College of
Electronic Engineering, Chongqing University of Post and 
Telecommunication, Chongqing 400065, China}
\author{Micheal F. Reid}
\affiliation{Department of Physics and Astronomy, 
University of Canterbury, Christchurch, New Zealand
}

\date{\today}
\begin{abstract}
Effective interaction operators usually act on a restricted model space
and give the same energies (for Hamiltonian) and matrix elements (for transition
operators etc.) as those of the original operators between the
corresponding true eigenstates. Various types of effective operators are 
possible. Those well defined effective operators have been shown being
related to each other by similarity transformation. Some of the
effective operators have been shown to have connected-diagram
expansions. It is shown in this paper that under a class of very general
similarity transformations, the connectivity is conserved. The
similarity transformation between hermitian and non-hermitian
Rayleigh-Schr\"{o}dinger perturbative effective operators is 
one of such transformation and hence the connectivity can be
deducted from each other.

\end{abstract}

\maketitle

\section{Introduction}

The full Hilbert space time-independent Hamiltonian H can be transformed
into an effective Hamiltonian $H_{\rm eff}$, which acts on a restricted
model space and gives the desirable exact eigenvalues.
\cite{Bra1967,Lin1985,Hurtubise3,Lin1987,Stolarczyk,KilJ2003a}
  Correspondingly,
effective transition operator $O_{\rm eff}$ is introduced to give the
same matrix elements while acting between the  model space eigenstates
as the original transition operator $O$ acting between the corresponding
true eigenstates.\cite{Bra1967,Hurtubise3,Brandow1,Suzuki} The explicit
forms of $H_{\rm eff}$ and $O_{\rm eff}$ 
are generally much more complicated than those of the original
Hamiltonian $H$ and operator $O$, which act on the infinite-dimensional
Hilbert space. Nonetheless, they are important and convenient {\it ab
initio} computation tools for a variety of problems. Another general
application of them is to give theoretical justification to
phenomenological Hamiltonian and transition operator,\cite{comment} such
as those used in $f^N$ energy level and transition intensity calculations.

$H_{\rm eff}$ and $O_{\rm eff}$  have been widely explored with
both perturbative methods (Brillouin-Wigner scheme,
Rayleigh-Schr\"{o}dinger and time-dependent scheme)
\cite{Bra1967,Lin1985,Goldstone,Hugenholtz},and
non-perturbative methods, such as 
iterative schemes and multi-reference open-shell coupled cluster
theories.\cite{Mukherjee,Lindgren2,Mukherjee1} The results were
initially single reference theory \cite{Goldstone,Hugenholtz} and have 
been generalized with many efforts to multi-reference cases for both
model space and Fock space.\cite{Bra1967,Lin1985,Stolarczyk,Mukherjee,Lindgren2,Mukherjeel_Pal,Jezioski_Paldus} 
 Well defined effective Hamiltonians and operators
for model space are related to the original operators by a similarity
transformation.\cite{Hurtubise3} Similarity transformation play very
important rules in the 
derivation of effective Hamiltonian and is assumed to take certain 
exponential (normal) forms in coupled-cluster methods, hence the
connectivity of effective operators follows trivially. In perturbative
methods, the projected transformation, 
i.e., the wave operator, is defined by order by order
expansions. Various effective operators can then be defined with the wave
operator and are related to each other by a class of similarity
transformation.\cite{Suzuki} In this paper, the connectivity is proved to be
conserved under such transformation. Therefore if one of those effective
operators has been proved  
to be connected, then the connectivity of all the others follows. 

\section{Formulation of the problem and lemmas}
Following Lindgren,\cite{Lin1985,Lindgren_JPB,Finley} the
effective multi-reference perturbative Hamiltonian for $H=H_0 + V$,
which produces a set 
of exact eigenvalues, is
\begin{equation}
H_{\rm eff}^{(0)} = PH\Omega P,
\end{equation}
where $P$ is the model space projector, $\Omega = 1 + \chi $ is the wave
operator, which produces exact eigenstates while acting on the model
function (eigenfunction of $H_{\rm eff}^{(0)}$), and $\chi$ has nonzero
matrix elements between the space $Q$ and $P$ only, where $Q$ is the
orthogonal space of $P$. The 
superscripts $(0)$ is used to distinguish this effective Hamiltonian from
others. Such superscripts is consistent with that of Suzuki {\it et
  al.}\cite{Suzuki}, which will be used throughout this paper.

The effective Hamiltonian $H^{(0)}_{\rm eff}$ is not hermitian and
therefore has 
different and non-orthonormal bra eigenfunctions
($~_b\Bra{\Phi^{\alpha}_0}$)and ket 
eigenfunctions ($\Ket{\Phi^{\beta}_0}_k$,
which can be bi-orthonormalized and are related to exact eigenstates of
$H$ with wave operator, i.e.,
\begin{eqnarray}
~_b\Bra{\Phi^{\alpha}_0} H_{\rm eff} \Ket{\Phi^{\beta}_0}_k 
&=& E_{\alpha} \delta_{\alpha\beta},
\label{eigen_eq}\\
~_b\Bra{\Phi^{\alpha}_0}\Ket{\Phi^{\beta}_0}_k 
&=& \delta_{\alpha\beta},
\label{biorth_eq}\\
   \Ket{\Phi^{\alpha}} 
&=& \Omega \Ket{\Phi^{\beta}_0}_k
\label{ket_eq}\\
   \Bra{\Phi^{\beta}}
&=& ~_b\Bra{\Phi^{\alpha}_0}
(\Omega^+\Omega)^{-1}\Omega^+.
\label{bra_eq}
\end{eqnarray}
The nonhermitian effective operator $O_{\rm eff}$ of operator $O$ for this
biorthonormal bases is 
\be
O_{\rm eff} = (\Omega^+\Omega)^{-1}\Omega^+ O \Omega,
\ee
which has been proved to have connected
diagrammatic expansion.\cite{paper1} The model space projector can be
written with the biorthonormal as
\be
P =\sum_{\alpha} \Ket{\Phi^{\alpha}_0}_k ~_b\Bra{\Phi^{\alpha}_0}.
\label{P_eq}
\ee

The hermitian effective Hamiltonian and associated hermitian
operator \cite{Bra1967,Brandow1,Bulaevski} are
\begin{eqnarray}
\label{Heff1/2}
H_{\rm eff}^{(-1/2)} 
&=& (\Omega^+\Omega)^{1/2} H^{(0)}_{\rm eff}
  (\Omega^+\Omega)^{-1/2} \nonumber\\
&=& H_0 + (\Omega^+\Omega)^{1/2} V\Omega,\\
O_{\rm eff}^{(-1/2)} 
&=& (\Omega^+\Omega)^{-1/2}\Omega^+ O \Omega
  (\Omega^+\Omega)^{-1/2}\nonumber\\
&=&  (\Omega^+\Omega)^{1/2} O^{(0)}_{\rm eff}
  (\Omega^+\Omega)^{-1/2},
\end{eqnarray}
where the second equality in (\ref{Heff1/2}) holds only for strictly
degenerate model space.

It can be seen that the hermitian effective Hamiltonian and operator are 
related to the nonhermitian ones by a similarity
transformation. We will show that the relations between the effective
Hamiltonian 
and operator and the original Hamiltonian and operator are
also similarity transformations (followed by a projection to model
space, which can be avoided). Define a transformation operator $T_n$
(arbitrary real $n$) as
\begin{equation}
T_n = (1+\chi - \chi^+)(1+\chi^+\chi + \chi\chi^+)^n.
\label{Tn_eq}
\end{equation}
As $\chi$ has only matrix elements between $Q$ and $P$, it can be shown
that
\be
\chi^2 = \chi^{+}~^2 = 0,
\label{chisqr}
\ee
and the $T_n^{-1}$ can be derived with these properties as
\be
T_n^{-1} = (1+\chi^+\chi + \chi\chi^+)^{-n-1} (1-\chi+\chi^+).
\label{Tmn_eq}
\ee
It can be seen that $T_{-1/2}$ is a hermitian transformation.
The similarity transformations of Hamiltonian, effective operator and
eigenstates generated by $T_n$ are
\begin{eqnarray}
\tilde{H}_n &=& T_n^{-1}HT_n,\\
\tilde{O}_n &=& T_n^{-1}OT_n,\\
\Ket{\Phi^{\alpha}} &=& T_n \Ket{\Phi^{\alpha}_n}_b,\\
\Bra{\Phi^{\beta}} &=& ~_b\Bra{\Phi^{\beta}_n} T_n^{-1},
\end{eqnarray}
where $\Phi^{\alpha}$'s and $\Phi^{\alpha}_n$'s are eigenstates for $H$
and $\tilde{H}_n$ respectively.
The decoupling condition 
\be
Q\tilde{H_n} P = 0,
\ee
is required to diagonalize the transformed  Hamiltonian in model
space. It can be shown that it is satisfied as follows:
\begin{widetext}
\begin{eqnarray}
    Q\tilde{H_n}P
&=& Q(1+\chi\chi^+)^{-n-1}
 (1 -\chi)H(1+\chi)P(1+\chi^+\chi)^n \\  
&=& Q(1+\chi\chi^+)^{-n-1} (1 -\chi)(\sum_{\alpha}E_{\alpha}
 \Ket{\Phi_{\alpha}}\Bra{\phi_{\alpha}^0} P(1+\chi^+\chi)^n\\
&=& Q(1+\chi\chi^+)^{-n-1}(1-\chi)(1+\chi)PH_{\rm eff}^{(0)}(1+\chi^+\chi)^n\\ 
&=& 0.
\end{eqnarray}
\end{widetext}
Furthermore,
\be
P\tilde{H}_nQ = (Q\tilde{H}_{-(n+1)}P)^+ = 0.
\ee
As there is no matrix element of $\tilde{H_n}$ between model space and
the orthogonal space, the diagonalization can be done in model space to
give exact eigenvalues and model functions. The effective Hamiltonian
for the model space can be simplified as
\begin{widetext}
\begin{eqnarray}
    H^{(n)}_{\rm eff} &=& P\tilde{H_n}P \\
&=& (P+\chi^+\chi)^{-n-1} (P+\chi^+)H(P+\chi)(P+\chi^+\chi)^n \\
&=& (P+\chi^+\chi)^{-n-1}(P+\chi^+)
    (\sum_{\alpha}(P+\chi)\Ket{\Phi^{\alpha}}_k~_b
    \Bra{\Phi^{\alpha}}_0E_{\alpha})
    (P+\chi^+\chi)^n \\
&=& (P+\chi^+\chi)^{-n-1}
(P+\chi^+)(P+\chi)H^{(0)}_{\rm eff}(P+\chi^+\chi)^n \\
&=& (P+\chi^+\chi)^{-n}H(P+\chi)(P+\chi^+\chi)^n.
\end{eqnarray}
\end{widetext}
It can be shown that both $T_n^{-1}\Ket{\Phi_{\alpha}}$ and
$\Bra{\Phi_{\alpha}}T_n$ are in model space, which are the ket and bra
model functions respectively. Therefore
the effective operator for model space can be derived by projecting
 $\tilde{O}_n$ to model space, i.e.,
\begin{eqnarray}
&&O_{\rm eff}^{(n)} = P\tilde{O_n}P\\\nonumber
&=&
(P+\chi^+\chi)^{-n-1}(P+\chi^+)O(P+\chi)(P+\chi^+\chi)^n. 
\end{eqnarray}
Such results have been derived by Suzuki and
Okamoto\cite{Suzuki} in other ways. They showed that the effective
Hamiltonian is related to the origin Hamiltonian by a similarity
transformation. However, the similarity transformation is not suitable
for the effective operator. Here a
similarity transformation for both Hamiltonians and operators have been
shown. This is what we have been expected, as Hamiltonian is only a special
operator which need be decoupled. Hereafter we do not distinguish
between them and the ``operator'' refers to both.

Various effective operators $\tilde{O}_n$  are related to the original
operator by similarity 
transformation, and therefore are related to each other by similarity
transformation, i.e.,
\begin{eqnarray}
\tilde{O}_{n+a} &=& (1+\chi^+\chi+\chi\chi^+)^{-a}
  \tilde{O}_n  (1+\chi^+\chi+\chi\chi^+)^{a},\\
O^{\rm eff}_{(n+a)} &=& (P+\chi^+\chi)^{-a}O^{\rm eff}_{(n)}
                      (P+\chi^+\chi)^{a}.
\end{eqnarray}
Such property between operators are very important, since
commutation relations, which are closely related to symmetries,
 are conserved under similarity transformation.\cite{Hurtubise3} 

It is well known that the similarity transformation generated by a
exponential function of a connected operator (referred as cluster
function) preserve the connectivity, which has been the bases of coupled 
cluster methods, i.e.:

{\it If $S$ and $O$ are connected, then $\exp(-S)O\exp(S)$ is connected.
}

The proof of this is straightforward by using the famous
Campbell-Baker-Hausdorff formulas.

We will show that the similarity transformation between various
perturbative effective operators generate by $(P+\chi^+\chi)^a$, or $\X
^a$, also preserves the connectivity, i.e.:

{\it  Theorem:  $\X ^a O \X ^{-a}$ is connected if $O$ is connected, where
  $\Omega$ is perturbative wave operator for complete multi-reference
  model space. The completeness means that model space contains all
  bases which can be formed by distribution the valence electrons among
  the valence shells.
}

 The following lemmas are used to prove this theorem

{\it Lemma 1}. The RS perturbative expansion of the wave operator
$\Omega$ can be written in a exponential form, i.e., 
\begin{equation}
\Omega = \norm{\exp(S)},
\end{equation}
where the curly brackets mean that the creation and annihilation
operators within them are rearranged into normal form with respect to a
closed-shell state. This notation for normal form will be used
throughout this paper. In the case of quasi-degenerate complete model
space, $S$ is a sum of connected diagrams.

{\it Lemma 2}. $\norm{\exp(S_1)} \norm{\exp(S_2)} =
\norm{\exp(S_{S_1S_2})}$, where $S_1$, 
$S_2$ and $S_{S_1S_2}$ are all connected. $S_{S_1S_2}$ is the connected
part of $\norm{\exp(S_1)}\norm{\exp(S_2)}$. 

{\it Lemma 3}. $xO-Ox = O^{(1)} + (1-\delta) x O^{(1)} + \delta O^{(1)} x$,
where $x = \norm{\exp(S)}-1$, $S$, $O$ and $O^{(1)}$  are 
connected and in normal form, and $\delta$ is an arbitrary real number. The
order of $O^{(1)}$ is higher than $O$ by at least one, where the order
is the smallest number of $V$ of all the terms of the operator concerned.

{\it Lemma 4}. Define $\alpha^{(m)(n)}_i$ (integer $n$ and $i$,
$n=1,2,\cdots,\infty$, $0\leq i \leq n$) recursively as
\begin{eqnarray}
\alpha^{(0)(n)}_i &=& \vtwo{a}{n-i}\vtwo{-a}{i},\\
\alpha^{(k)(n)}_i &=& \sum\limits_{j=0}^i[\alpha^{(k-1)(n+1)}_j +
\alpha^{(k-1)(n)}_j],
\end{eqnarray}
where $k = 1,2, \cdots$.
The following equality holds for arbitrary positive integer $m$ and $n$:
\be
\sumb\alpha^{(m)(n)}_i = 0.
\label{lemma4}
\ee

In addition to applying the theorem to show the connectivity of various
MBPT effective Hamiltonians and transition operators, the theorem and lemmas
can also be used to show the connectivity of various effective Hamiltonian
in Coupled-Cluster (CC) theories\cite{Lin1987,Sta1993,Noo1996,Noo2000}.
In those theories various similarity transformations, generated by 
\begin{equation}
T= \{ \exp \hat S \}
\end{equation}
to the right and $T^{-1}$ ( generally $\not= \{\exp (-\hat S)\}$ to the left,
have been used to transform original Hamiltonian operator $H$ or CCSD
 (CC Singleton and doubleton excited contribution) Hamiltonian operator
 $\exp({-T_1-T_2}) H \exp({T_1+T_2})$ into Coupled-Cluster 
 effective Hamiltonian which have certain zero components convenient for
 calculation of eigenvalues and eigenvectors. From the theorem and Lemma 2
 it is straightforward to show that all such transformations preserve 
 connectivity. Moreover, if necessary, more general similarity transform 
 generated by $T^{a}$ ($a$ an arbitrary number) can be used in CC methods.
 
\section{Proof of the theorem and the lemmas}
Lemma 1 has been proved by Lindgren by using factorization theorem and
mathematical induction \cite{Lin1985,Lindgren_JPB},
and Lemma 2 has been proved in another paper\cite{paper1}.
We shall prove the theorem to be true firstly by using these lemmas and
then prove lemma 3 and lemma 4 afterwards.

\subsection{Proof of the theorem}

Denoting  $x=\Omega^+\Omega -1$ and using the definition of
$\alpha^{(0)(n)}_i$ in lemma 4, we have 
\begin{eqnarray}
&&(\Omega^+\Omega)^aO(\Omega^+\Omega)^{-a}
= O + \suma \sumb \alpha^{(0)(n)}_i x^{n-i} O x^i\nonumber\\
&&= \suma\sumb \beta^{(0)(n)}_i x ^{n-1-i}(xO-Ox)x^i,
\label{atob}
\end{eqnarray}
where
\be
\beta^{(m)(n)}_i = \sum\limits_{j=0}^{i}\alpha^{(m)(n)}_j,~(i=0,1,\cdots,n),
\label{bmni}
\ee
and the condition $\beta^{(0)(n)}_n=0$, which follows from lemma 4, has
been used in deriving equality (\ref{atob}).
It can be seen from lemma 1 and lemma 2 that the $x$ in (\ref{atob}) can 
be written as 
\be
x = \{\exp(S)\}-1,
\ee
where $S$ is the connected part of $\exp(S_1)\exp(S_2)$ that contain only
valence creation and annihilation operators. Applying the $\delta=0$ case
of lemma 3  to (\ref{atob}), we get
\begin{widetext}
\begin{eqnarray}
(\Omega^+\Omega)^aO(\Omega^+\Omega)^{-a}
&=& O+\beta^{(0)(1)}_0 O^{(1)}  
  + \suma\sumb(\beta^{(0)(n+1)}_i + \beta^{(0)(n)}_i) x^{n-i}O^{(1)}x^i
\\
&=& O+\alpha^{(0)(1)}_0 O^{(1)}  
  + \suma\sumb\alpha^{(1)(n)}_ix^{n-i}O^{(1)}x^i.
\end{eqnarray}
\end{widetext}
As shown by lemma 4 that $\sumb \alpha^{(1)(n)}_i =0$. We can
simply repeat the above procedure to
arbitrary $m$ and get
\begin{eqnarray}
&&   (\Omega^+\Omega)^aO(\Omega^+\Omega)^{-a}
 = O \nonumber\\
 &&+\sum\limits^{m-1}_{l=0} b^{(l)(1)}_0 O^{(l)}
    +\suma\sumb \alpha^{(m)(n)}_i x^{n-i}O^{(m)}x^i.
\label{exp_2}
\end{eqnarray}
We conclude by mathematic induction that Eq. \ref{exp_2}
holds for arbitrary $m>0$. As  $O^{(m)}$ is connected and its order
increases by at least 1 as $m$ increases by 1, we have proved that to
arbitrary large but finite order, the expansion of $(1+x)^{a}O(1+x)^{-a}$
is connected. 

\subsection{Proof of lemma 3}
A special case of the lemma 4 of \cite{paper1} is
\begin{eqnarray}
    \norm{\exp(S)}O
&=& \norm{\exp(S)O^L}\\
&=&  \norm{\exp(S)O} +\norm{\exp(S) {O}^L_1}
\label{lemma4_L}
\end{eqnarray}
where $O^L$ is the connected part of
$\{\exp(S)\}O$, and ${O}^L_1=O^L_1-O$ is the connected part of
$(\{\exp(S)\}-1)O$, whose order is higher than $O$ by at least one.

Denoting $x=\{\exp(S)\}-1$, we have
\be
xO = \norm{xO}+\norm{(1+x)O_1^L}.
\ee
Similarly, it can be proved that
\be
Ox=\norm{Ox}+\norm{O_1^R(1+x)},
\ee
where $O_1^R$ is the connected part of Ox, whose order is also higher
than $O$ by at least 1. 

The lemma 5 of \cite{paper1}, which has been proved, is
\begin{equation}
\norm{\exp(S)O} =\norm{\exp(S)}O\prime,
\label{lemma5}
\end{equation}
where $S$, $O$ and $O\prime$ are all connected, and the order of
$O\prime$ is the same as $O$.

The following equation can be derived using the latest three equations 
\begin{eqnarray}
xO-Ox &=& \norm{(1+x)(O_1^L-O_1^R)}\nonumber\\
      &=& (1+x)O^{(1)},\label{delta=0}
\end{eqnarray}
where $O^{(1)}$ is connected and the order is no less than $O_1^L-O_1^R$.

The case $\delta \neq 0$ can also be proved with some mathematical
manipulation. Note that $O^{(1)}$ depends on the value of $\delta$.

\subsection{Proof of Lemma 4}
The case $m=0$ of Eq.\ \ref{lemma4} can be proved directly by the
binomial expansion of $(1+x)^a(1+x)^{-a}$. 

For $m=1$, we have
\begin{eqnarray}
&&\alpha^{(1)(n)}_i
= \sum\limits_{i_1=0}^{i}
    [\alpha^{(0)(n+1)}_{i_1} + \alpha^{(0)(n)}_{i_1}] \nonumber\\
&&= \sum\limits_{i_1=0}^{i}
    \left [\vtwo{a}{n+1-i_1} \vtwo{-a}{i_1}
          +\vtwo{a}{n-i_1}\vtwo{-a}{i_1}
    \right ]\nonumber\\
&&= \sum\limits_{i_1=0}^{i}
    \vtwo{a}{n-i_1}\vtwo{-a}{i_1}
    \left ( 1 + \frac{1+a-(n+1-j)}{n+1-j} \right )\nonumber\\
&&= \sum\limits_{i_1=0}^{i}
    \vtwo{1+a} {n+1-i_1}\vtwo{-a}{i_1}.
\end{eqnarray}
Suppose, for a given $k$, that the
following equation holds:
\be
\alpha^{(k)(n)}_i = 
 \sum\limits_{i_1=0}^i \sum\limits_{i_2=0}^{i_1}
   \cdots \sum\limits_{i_k = 0}^{i_{k-1}} 
  \vtwo{a+k}{n+k-i_k}\vtwo{-a}{i_k}.
\label{akni_final}
\ee

The recursive relation tells us that
\begin{widetext}
\begin{eqnarray}
\alpha^{(k+1)(n)}_j
 &=& \sum\limits_{i=0}^{j}( \alpha^{(k)(n+1)}_i + \alpha^{(k)(n)}_i)
     \nonumber\\ 
 &=& \sum\limits_{i=0}^{j}\sum\limits_{i_1=0}^{i} \cdots
      \sum\limits_{i_k = 0}^{i_{k-1}}
      \vtwo{a+k}{n+k-i_k}\vtwo{-a}{i_k}
       \left (
        \frac{a+k+1-(n+1+k-i_k)}{n+1+k-i_k}+1
       \right ) \nonumber\\
 &=& \sum\limits_{i=0}^{j}\sum\limits_{i_1=0}^{i} \cdots
      \sum\limits_{i_k = 0}^{i_{k-1}}
      \vtwo{a+k+1}{n+k+1-i_k}\vtwo{-a}{i_k}
\end{eqnarray}
\end{widetext}
By mathematical induction, we conclude that Eq.\ \ref{akni_final} holds
for all $k$. 
Then Eq.\ \ref{lemma4} reduces to the following equation:
\begin{eqnarray}
&&\sumb\alpha^{(m)(n)}_i 
= \sumb \sum\limits_{i_1=0}^{i}\cdots \sum\limits_{i_m = 0}^{i_{m-1}}
    \vtwo{a+m}{n+m-i_m}\vtwo {-a}{i_m} \nonumber\\
&&= \sum\limits_{i_m = 0}^n 
    \left [
      \vtwo{a+m}{n+m-i_m}\vtwo {-a}{i_m}
      \sum\limits_{i_{m-1}=i_m}^n\cdots \sum\limits_{i_1=i_2}^n
       \sum\limits_{i=i_1}^n 1
     \right ].
\label{sum_amni}
\end{eqnarray}
It is straight forward to prove by mathematical induction that
\be
 \sum\limits_{i_{m-1}=i_m}^n\cdots \sum\limits_{i_1=i_2}^n
   \sum\limits_{i=i_1}^n 1
 = \frac{(n+m-i_m)!}{m! (n-i_m)!}.
\ee
Substituting the corresponding summations in Eq.\ \ref{sum_amni} with this 
result, we get 
\begin{eqnarray}
\sumb\alpha^{(m)(n)}_i 
&=& \sum\limits_{i_m=0}^n
    \vtwo{a+m}{n+m-i_m}\vtwo {-a}{i_m}
    \frac{(n+m-i_m)!}{m!(n-i_m)!}\nonumber\\
&=& \sum\limits_{i_m=0}^n
    \vtwo{a+m}{m}
    \vtwo{a}{n-i_m}
    \vtwo{-a}{i_m}\\
&=& \vtwo{a+m}{m}\delta_{n0}.
\end{eqnarray}

\section{Conclusion}
It has been shown that the perturbative effective Hamiltonian and operator
are related to the original Hamiltonian and operator respectively by the
same similarity transformation, which includes the hermitian special
case. Such transformation conserves the commutation relations and hence most
symmetry properties. Various effective Hamiltonians and effective
operators respectively are related to each other by a similarity
transformation generated by $\{\exp S\}^a$, where $S$ is connected, $a$
is an arbitrary real number, and curved 
bracket means normal form. An effective Hamiltonian or operator with
connected-diagram expansion will be transformed into a new operator with
connected-diagram expansion, consequently the connectivity can be
deduced from each other. In particular,The hermitian effective Hamiltonian and
operator are related to the simplest non-hermitian
effective Hamiltonian and operator respectively by such a
transformation, and therefore are connected from the fact
that the later effective Hamiltonian and operator has been proved to be
connected.\cite{paper1} This rigorous mathematic proof saves one from
understanding the complicated demonstration by recursive insertion of
energy diagrams\cite{Bra1967}.

\end{document}